\documentclass[preprint]{JHEP3} % 10pt is ignored!

\usepackage{epsfig,multicol}

\def\beq{\begin{equation}}

\def\eeq{\end{equation}}
\def\eq#1{{eq.~(\ref{#1})}}

\parskip=\itemsep               %?

\newcommand{\beqar}[1]{\begin{eqnarray}\label{#1}}

\newcommand{\eeqar}{\end{eqnarray}}

\def\thefootnote{\fnsymbol{footnote}} 

\title{\Large  \bf More Remarks on High Energy Evolution.}
\author{A. Kovner and M. Lublinsky \\
Physics Department, University of Connecticut, \\ 2152 Hillside
Road, Storrs, CT 06269-3046, USA}

\abstract{We discuss several issues related to recent work on high energy evolution. In particular we show that the Hilbert space for action of the operator of the evolution can be conveniently defined by the space of weight functional introduced recently in \cite{kl4}. These weight functionals automatically implement the correct properties of the charge density correlators, thus eliminating the need in explicit introduction of the Wess Zumino term. We also discuss various aspects of Dense Dilute Duality in the toy dipole model.}

\begin{document}
%\noindent

%\begin{flushright}

%\parbox[t]{10em}{

% \today 

%\end{flushright}

%\vspace{1cm}

%\begin{center}

%\end{center}

%*********************************************************************************  

%*********************************************************************************  

\def\thefootnote{\arabic{footnote}} 

\section{Introduction}

The ideas of gluon saturation\cite{GLR} have been vigorously pursued in the last ten years or so \cite{Mueller,mv,reggeon,balitsky,JIMWLK,cgc,Kovchegov}.
The result of these efforts is the set of evolution equations which describes the change of the scattering amplitude at high energy in the regime where the hadronic wave function resembles a dense gluonic medium - the so called JIMWLK equation\cite{balitsky,JIMWLK,cgc,Kovchegov} (for reviews see \cite{IV,Weigert,JK,zakopane}).

The latest developments in this dynamic field are related to what is sometimes called Pomeron loops, 
or Fluctuations or alternatively Saturation effects in the projectile wave function
\cite{IM,kozlov,LL1,shoshi1,IMM,balitsky1,janik,
LL2,IT,MSW,LL3,kl,kl1,something,blaizot,kl4,genya,
Braun05,SMITH,MMSW,HIMS,Balitsky05,Kovchegov05}. The corrections due to these effects 
clearly must be important for the preasymptotic regime, but should also affect the more exclusive observables at very high energy. The last year or so has seen a lot of activity in this direction. 
Much has been understood during the last year, but many questions remain open.
In particular the structure of the "Hilbert space" on which the evolution kernel acts \cite{kl} as well as the precise physical meaning of the Dense-Dilute Duality (DDD) introduced in \cite{something} are being actively debated \cite{kl4,SMITH,MMSW,HIMS,Balitsky05}.

This paper presents several remarks on these issues. The remarks are not necessarily causally or logically connected, but they all pertain to the questions mentioned above.

As regards the first issue, one of the questions is the following. The evolution is represented usually as acting on the weight function $W[\rho(x,x^-)]$, where the necessity of introducing the longitudinal coordinate $x^-$ stems from
the desire to represent quantum noncommuting operators $\hat\rho^a(x)$ in terms of commuting "classical fields". It was shown in \cite{kl} that this is achievable by endowing the quantum operator with an additional "ordering" variable. One also needs the weight functional $W$ to ensure that the correlators of  the type $\langle\rho(x_1^-)...\rho(x_n)^-\rangle=\int D[\rho]W[\rho]\rho(x_1^-)...\rho(x_n^-)$ depend only on the ordering of the variables $x^-_i$ rather than on their values, and that an exchange of two such variables must lead to the appearance of the $SU(N)$ structure constant. Namely if in the correlation function the two values of the ordering variable $x^-_i,x^-_j$are adjacent to each other and $x^-_i>x^-_j$, the following property must hold 
\begin{equation}
\int D[\rho]W[\rho]...\rho^a(x_i^-)\rho^b(x_j^-)...=\int D[\rho]W[\rho]...\rho^a(x_j^-)\rho^b(x_i^-)...
+if^{abc}\int D[\rho]W[\rho]...\rho^c(x_i^-)...
\label{property}
\end{equation}

It was also shown in \cite{kl} that these requirements can be satisfied by a "weight function" $W$ which contains 
a  Wess-Zumino term. The explicit representation of such a Wess - Zumino term is awkward, since it has to be local in transverse space and has to allow for an arbitrary representation of the colour charge at every transverse position\cite{kl}. 

On the other hand in \cite{kl4} we have introduced a different representation of the weight functional
\begin{equation}
W[\rho^a]\,=\,\Sigma[R]\,\delta[\rho(x,x^-)]
\label{hilbert}
\end{equation}
where 
\begin{equation}
R^{ab}(x)\,=\,\left[P\exp\{\int dx^-T^c{\delta\over\delta\rho^c(x,x^-)}\}\right]^{ab}
\end{equation}
and $\Sigma$ - an arbitrary functional. A similar representation was introduced earlier in the framework of the dipole model\cite{IM}, with the crucial difference that the exponential in functional derivatives was truncated at a low order in the Taylor expansion. 
Although the formula eq.(\ref{hilbert}) has a very natural interpretation from the point of view of partonic content of the wave function, the question whether it satisfies the property eq.(\ref{property}) has not been addressed before.
Our observation is that the Hilbert space for the action of the high energy evolution kernel 
is indeed naturally given by the vector space eq.(\ref{hilbert}). In particular we show that
the necessary condition eq.(\ref{property}) is satisfied by any element of the space defined by eq.(\ref{hilbert}).
We thus note that this form constitutes a simple alternative to the Wess-Zumino term as far as the correct representation of the charge density commutation relations is concerned. This discussion along with some additional remarks which are mostly meant to clarify and explain in more detail our earlier results  \cite{kl,kl1,something,kl4}
is contained in Section 3. 

In Section 4 we discuss a dipole toy model with no transverse coordinates. We ask the question what is the DDD transformation
and what is the analogue of the Seld-Duality condition in terms of the dipole degrees of freedom.
By applying the DDD transformation we construct the dual to the Mueller`s dipole evolution.
In the framework of this model we explicitly construct a self dual evolution which is valid both in the dilute and dense limits. In the intermediate regime the model is incomplete and misses correct treatment of multiple dipole emissions. This is reminiscent of the self dual variant of the high energy evolution suggested in \cite{SMITH} and discussed recently in \cite{Balitsky05}. The value of this discussion is mainly in the simplicity of the model and transparency of the mathematics involved which thus allows us to clearly see some features of the evolution which are much more difficult in real QCD.

We start in Section 2 by a short review of the elements we would like to discuss.

\section{JIMWLK and other letters }

We start by describing the general setting of the high energy evolution.
We consider the leftmoving projectile with the wave function $|P\rangle$ colliding with the rightmoving target with wave function $|T\rangle$. The total rapidity of the process is $Y$, of which the target carries rapidity $Y_0$. 
The $S$-matrix for the process is given by
\begin{equation}
{\cal S}(Y)\,=\,\int\, D\alpha_T^{a}[\tau]\,\, W^T_{Y_0}[\alpha_T(x,\tau)]\,\,\Sigma^P_{Y-Y_0}[\alpha_T(x,\tau)]\,.
\label{ss}
\end{equation}
Here $\alpha$ denotes the $gA^-$ component of the vector potential due to the target charge density $\rho_T$.  It defines the scattering matrix of a single  gluon projectile at transverse position $x$ on the target, which in the eikonal approximation is given by\footnote{In our convention
the variable $\tau$ is rescaled to run from 0 to 1. 
Thus $\tau=0$ corresponds to $x^-=-\infty$ 
while $\tau=1$ to $x^-=+\infty$.}
\begin{eqnarray}\label{S}
S(x;0,\tau)\,\,=\,\,{\cal P}\,\exp\{i\int_{0}^{\tau} d\tau' T^a\alpha^a_T(x,\tau')\}\,.
\end{eqnarray}
where $T^a_{bc}=if^{abc}$ is the generator of the $SU(N)$ group in the adjoint representation.
The field $\alpha_T$
obeys the classical equation of motion  \cite{JIMWLK,cgc} 
\beq
\alpha^a_T(x,\tau)T^a\,\,=g^2\,\,{1\over \partial^2}(x-y)\,
\left\{S^\dagger(y;0,\tau)\,\,\rho^{-a}_T(y,\tau)\,T^a\,\,S(y;0,\tau)\right\}
\eeq
The function $\Sigma^P$ in eq.(\ref{ss})
 is the eikonal scattering matrix for a composite projectile averaged over the projectile wave function, see \cite{something}
\begin{equation}
\Sigma^P[\alpha_T]\,\,=\,\,\int d\rho_P\,\,W^P[\rho_P]\,\,
\,\exp\left\{i\int_{0}^{1} d\tau\int d^2x\,\rho_P^a(x,\tau)\,\alpha_T^a(x,\tau)\right\}\label{s}
\end{equation}
with $x_i$ - the transverse coordinates of the gluons in the projectile. $W^P$ is the "weight functional"
which determines the distribution of the charge density in the projectile.

The evolution of the $S$-matrix with energy is given by the general expression of the form:
\begin{equation}
\frac{d}{d\,Y}\,{\cal S}\,=\,\int\, D\alpha_T^{a}\,\, W^T_{Y_0}[\alpha_T(x,\tau)]\,\,\,
\chi\left[\alpha_T,\frac{\delta}{\delta\,\alpha_T}\right]\,\,\,
\Sigma^P_{Y-Y_0}[\alpha_T(x,\tau)]\,.
\label{hee}
\end{equation}
The kernel of the evolution $\chi$ can be viewed as acting
either to the right on $\Sigma$ or to the left on $W$:
\begin{equation}
{\partial\over\partial Y}\,\Sigma^P\,\,=\,\,\chi\left[\alpha_T,\,{\delta\over\delta\alpha_T}\right]\,\,\Sigma^P[\alpha_T]\,;
\ \ \ \ \ \ \ \ \ 
{\partial\over\partial Y}\,W^T\,\,=\,\,
\chi^\dagger\left[\alpha_T,\,{\delta\over\delta\alpha_T}\right]\,\,W^T[\alpha_T]\,.
\label{dsigma}
\end{equation}
 
Note that according to eq.(\ref{s}), the projectile-averaged $S$-matrix $\Sigma^P$ is the functional Fourier transform of the projectile weight functional $W^P$. Thus $\alpha_T$ plays the role of conjugate to $\rho_P$.
 The evolution of $\Sigma^P$ determines the evolution of its Fourier image $W^P$:
\begin{equation}
{\partial\over\partial Y}\,W^P\,\,=\,\,\chi\left[-i{\delta\over\delta\rho_P},i\rho_P\right]\,\,W^P[\rho_P]\,.
\label{dwp}
\end{equation}
Since the functional form of the evolution of the target and projectile weight functionals should be the same (projectile-target
democracy - PTD) comparing eq.(\ref{dwp}) with eq.(\ref{dsigma}) we conclude that the kernel $\chi$ must be selfdual  
\begin{equation}
\chi\left[\alpha,\,{\delta\over\delta\alpha}\right]\,=\,\chi\left[-i{\delta\over\delta\rho},i\rho\right]\,.
\label{sd}
\end{equation}
where now $\alpha$ and $\rho$ both refer to the same object, either target or projectile.
Thus if the target and wave function are treated on equal footing, the kernel $\chi$ must be invariant under the Dense Dilute Duality (DDD) transformation
\cite{something}:
\begin{equation}
\rho(x,x^-)\rightarrow -i{\delta\over\delta\alpha(x,x^-)}; \ \ \ \ \ \ \ {\delta\over\delta\rho(x,x^-)}\rightarrow i\alpha(x,x^-)
\label{duality}
\end{equation}

So far the approximations used in the literature to derive the kernel do not treat the target and the projectile equally. The approach \cite{JIMWLK,cgc} assumes that the the color charge density in the target is parametrically large ($\rho^a=O(1/\alpha_s)$). This yields the 
 the so called JIMWLK evolution operator \cite{JIMWLK,cgc}
\begin{eqnarray}\label{JIMWLK}
\chi^{JIMWLK}&&=\,
\frac{\alpha_s}{2\pi^2}\,\int_{x,y,z} {(z-x)_i\,(z-y)_i
\over (z-x)^2\,(z-y)^2} \,\,\times  \\
&&\left\{ {\delta\over \delta \alpha^a(x,0)}{\delta\over \delta \alpha^a(y,0)}\,+\, 
{\delta\over \delta \alpha^a(x,1)}{\delta\over \delta \alpha^a(y,1)}
\,- \,2\,{\delta\over \delta \alpha^a(x,0)}\,S(z;0,1)^{ab}\,{\delta\over 
\delta \alpha^b(y,1)}\right\}\,. \nonumber 
\end{eqnarray}
In this expression the functional derivative in \eq{JIMWLK} do not act on $S$ in the kernel, 
but on external 
factors only.
The same result follows from the calculation \cite{balitsky} which assumes that the projectile carries only small color charge density.
The derivation of \cite{kl} on the other hand assumes that the target is dilute (small charge density). This calculation yields
\begin{eqnarray}\label{KL}
\chi^{KLWMIJ}&&=\,-\,
\frac{\alpha_s}{2\pi^2}\int_{x,y,z} {(z-x)_i(z-y)_i
\over (z-x)^2(z-y)^2}\,\,\times  \\
&&\left\{\rho^a(x,0)\rho^a(y,0)+  \rho^a(x,1)\,\rho^a(y,1)-2\,  \rho^a(x,0)R(z,0,1)^{ab}\,\rho^b(y,1)\right\}\,.\nonumber
\end{eqnarray}
where the "dual Wilson line" $R$ is defined as
\begin{equation}
R(z;0,x^-)^{ab}=
\left[{\cal P}e^{\int_{0}^{x^-} d z^-T^c{\delta\over\delta\rho^c(z,\,z^-)}}\right]^{ab}
\label{rr}
\end{equation}
Just like in eq.(\ref{JIMWLK}) the functional derivatives in $R$  act only on external factors
 and not on the explicit factors $\rho^a(0)$ and $\rho^a(1)$ in Eq. (\ref{KL}).
Neither one of these expressions is self dual, rather they are related by the DDD transformation eq.(\ref{duality}).

Note that $R$ acts as the finite shift operator. When acting on a function of $\rho$ it shifts the charge density by the charge density of a single gluon 
\begin{equation}\label{rs}
R(z,0,1)\,\Sigma^P[\rho^a(x,x^-)]\,=\,\Sigma^P[\rho^a(x,x^-)+T^a\delta^2(x-z)]\,\, .
\end{equation}

More recently a step forward has been made by resumming part of the corrections due to
finite density effects in the wave function of a more dilute object (projectile in the case of JIMWLK, and target in the case of KLWMIJ)\cite{kl1,something,SMITH,Balitsky05}.
The reulting improved expressions are

\begin{eqnarray}
\chi^{JIMWLK+}&=&
\frac{1}{2\pi}\int_z\left\{ b^a_i(z,0,[{\delta\over \delta \alpha}])\,
b^a_i(z,0,[{\delta\over \delta \alpha}])\,+\, 
 b^a_i(z,1,[{\delta\over \delta \alpha}])\,b^a_i(z,1,[{\delta\over \delta \alpha}])
\right.\nonumber\\
&-&2\, \left. b^a_i(z,0,[{\delta\over \delta \alpha}])\,S^{ab}(z,0,1)\,b^b_i(z,1,[{\delta\over 
\delta \alpha}])\right\}
\label{notlarge}
\end{eqnarray}
\begin{eqnarray}\label{notsmall}
\chi^{KLWMIJ+}=&-&
\frac{1}{2\pi}\int_{z} \{ b^a_i(z,0,[\rho])\,b^a_i(z,0,[\rho])\,
+\,b^a(z,0,[\rho])\,b^a(z,0,[\rho])
\nonumber \\
&-& 2\,b^a_i(z,0,[\rho])\,R^{ab}(z,0,1)\,b^b_i(z,1,[\rho])\}\,.
\end{eqnarray}
The field $b^a_i$ satisfies the "classical" equation of motion
\begin{eqnarray}\label{b}
&&\left\{\partial_i+gf^{abc}b^c_i(z,[\rho])\right\}\partial^+ b_i^b(z,[\rho])=g\rho^a(z,x^-)\,;\nonumber\\
&&\epsilon_{ij}[\partial_ib^a_j(z,[\rho])-\partial_jb^a_i(z,[\rho])+gf^{abc}b^b_i(z,[\rho])b^c_j(z,[\rho])]=0\,.
\end{eqnarray}
Expressions eqs.(\ref{notlarge}) and (\ref{notsmall}) are again related by the
DDD transformation eq.(\ref{duality}). 

The KLWMIJ evolution equation was derived in \cite{kl} by directly considering the evolution of the projectile wave function.
In this framework the charge density operator depends only on the transverse coordinates $\hat\rho^a(x)$. As quantum operators the charge densities at the same transverse coordinate do not commute
\begin{equation}
[\hat\rho^a(x),\hat\rho^b(y)]=if^{abc}\hat\rho^c(x)\delta^2(x-y)
\label{com}
\end{equation}
Thus in general the averages of powers of $\hat\rho$ can not be represented simply by 
a functional integral with a weight functional which depends on the classical field $\rho^a(x)$ only. This is unlike JIMWLK, where such a representation is possible.
It is however possible to have a functional integral representation if the "classical field" is endowed with an additional "ordering" coordinate
\beq\label{wf}
\langle \hat\rho^{a_1}(x_1)... \hat\rho^{a_n}(x_n)\rangle \,\equiv\,
\int d\rho(x,x^-)\, \rho^{a_1}(x_1,x_1^-)... \rho^{a_n}(x_n,x_n^-)\,\,W^T[\rho(x,x^-)]
\eeq
On the right hand side of this expression the values of the additional ordering coordinate inherit the ordering of the operator factors on the left hand side: $x^-_1>x^-_2>...>x^-_n$. 
Clearly the weight functional $W$ can not be arbitrary. It must ensure two basic properties of the "classical" correlation functions. First, the correlation
function can not depend on the values of the coordinates $x^-_i$, but only on their relative ordering. Thus the correlation function should be a piecewise constant function of $x^-_i$. Second, under the interchange of the two adjacent values of $x^-_i$ the correlation function must emulate the effect of the commutator eq.(\ref{com})
\begin{equation}
\int D[\rho]W[\rho]...\rho^a(x_i^-)\rho^b(x_j^-)...=\int D[\rho]W[\rho]...\rho^a(x_j^-)\rho^b(x_i^-)...
+if^{abc}\int D[\rho]W[\rho]...\rho^c(x_i^-)...
\end{equation}
In particular this last property means that $W$ can not be real, and thus cannot be given the meaning of probability density. It was pointed out in \cite{kl} that a similar problem has appeared in the literature before \cite{polyakov} where it was shown that these properties are satisfied by a weight functional which contains an exponential of a Wess-Zumino term.

\section{On the operator ordering, $x^-$ vs $x^+$ and the Wess-Zumino-less weight functional.}
In this section we make several remarks on the structure of the evolution described in Section 2.

\subsection{On intimate relation between target and projectile}

A somewhat curious feature of the JIMWLK equation which is not frequently stressed in the literature, is that the same equation arises as the result of two apparently unrelated derivations. The original 
derivation \cite{JIMWLK,cgc}
was given directly for the target wave function (equation for $W^T$ in \eq{dsigma}), 
assuming explicitly that the target is dense, namely that the average of the single gluon $S$ matrix is close to zero. Nothing has been explicitly assumed there about the properties of the projectile. On the other hand, one can deduce the kernel
$\chi^{JIMWLK}$ in a different way assuming  the projectile is small and deriving the evolution of the 
projectile wave function \cite{MuJIM,kl1}. This then yields the evolution equation for the target $W_T$ via the functional Fourier transform, cf. eq.(\eq{dsigma}). In this derivation we apparently do not assume anything about the properties of the target, instead we have assumed that the projectile is dilute. Nevertheless the two ways lead to the same evolution equation for $W_T$. 
The question is how come one obtains the same equations under seemingly unrelated conditions? 

The answer is that these two conditions - diluteness
of the projectile and dense nature of the target are actually inextricably related through the form of the $S$-matrix 
eq.(\ref{s}). The JIMWLK kernel gives a correct approximation for the full kernel of the evolution of $W_T$ in the limit when $\alpha_T$
is large, or equivalently when $S$ is close to zero. On the other hand the evolution  of ${\cal S}$ eq.(\ref{hee}) is given by the functional
integral  over all $\alpha_T$. For $W^T$ which is sharply peaked around $S=0$, the integral picks out the large $\alpha_T$ region of $\Sigma^P[\alpha_T]$. 
Thus the approximation is safest when $\Sigma^P[\alpha_T]$ is also peaked around large $\alpha_T$.

This however is precisely the condition for the projectile to be dilute. This
is simply seen from \eq{s}. Mathematically 
it is the general nature of Fourier transformation that  $\Sigma^P[\alpha_T]$ is  peaked  
at large $\alpha_T$ if its Fourier image $W^P$ is dominated by small values of
$\rho_P$. 
The diluteness of the projectile is simply the statement that the projectile wave function contains only Fock space components with small number of gluons. 
Parametrically, if $\alpha_T$ is of order one as is assumed in the derivation of \cite{JIMWLK,cgc}, the projectile charge density that is picked out by the Fourier transform
is of order $O(1)$ also, which corresponds in our normalization to $O(1)$ number of particles in the projectile.
In this case the projectile averaged $S$-matrix $\Sigma^P$ has only
several first terms in the Taylor expansion in powers of  $S$, as each projectile gluon contributes a factor of $S$. Thus for a dilute projectile, the Taylor series expansion of $\Sigma^P$ has only small number of terms and is indeed dominated by $S\sim 0$.
 
The diluteness of projectile, and the dense nature
of the target therefore  go hand in hand. The corollary to this observation is that if one wishes to discuss scattering in the situation where the projectile is not dilute, the JIMWLK approximation to the target weight function is not necessarily applicable even if the target is dense. The projectile averaged $S$ matrix $\Sigma^P$ is then itself peaked around $S\sim 1$, and so even if the target is mostly dense, the tail of the distribution in $W^T$ at $\alpha_T\ll 1/g$ will contribute significantly to the
total cross section. Physically this is very reasonable. If the projectile is dense, it has large probability of scattering on a variety of configurations in the target wave function: the ones with large charge density as well as the ones with low charge density. Thus the tail of the distribution $W_T$ which determines the weight of dilute configurations in the target wave function  becomes important. 
A discussion of the evolution of the total scattering amplitude in this situation is to some extent superfluous, as it vanishes at all energies. However there are other less inclusive quantities which we believe will be strongly affected by this tail of the distribution. An example of such a quantity is diffractive scattering which is sensitive to the dilute peripheral region of the target. The  single gluon emission spectrum also may get sizeable contributions from the target configurations in the tail of the distribution.

\subsection{$x^+$ or $x^-$?}

Our second remark is essentially semantic. However since there has been some lack of clarity on this point in the recent literature, we thought it worth a comment.  

The issue in question is whether to call the ordering variable entering the KLWMIJ kernel $x^-$ or $x^+$. We prefer simply to think of it as of an ordering variable and clearly in practical terms its name does not matter. There is however a certain physics picture that suggests its analogy with $x^-$.
In particular consider the KLWMIJ+ equation eq.(\ref{notsmall}). It involves the factors depending on the "classical" field $b_i(x,\tau)$, which satisfies an equation of motion identical to the classical Yang-Mills equation in the presence of the source $\rho(x,\tau)$, where $\tau$ is the ordering variable. In our wave function approach \cite{kl} this equation arises in the following way. To begin with one has to solve operator equations for the {\it operator} $\hat b_i(x)$ which has an {\it operator } source $\hat\rho(x)$  without any $\tau$-dependence.  Solving these equations, for example in the perturbation theory, one has to keep track of the ordering of operators $\hat\rho^a$ as explained in the previous section. This procedure is streamlined by introducing the ordering variable $\tau$ and treating all fields as classical. The resulting "classical" equation is identical to the classical equation solved in \cite{first} if one identifies the ordering coordinate $\tau$ with the longitudinal spatial coordinate $x^-$. It is actually more consistent to interpret the ordering variable as the rapidity $\ln x^-/x^-_0$ rather than the longitudinal coordinate itself, since it is dimensionless, but we will keep referring to it $x^-$ for simplicity. This identification is natural in the sense that $x^-$ and not $x^+$ is the spatial coordinate of the target. As the whole action of KLWMIJ takes place in the projectile Schroedinger (light cone) wave function there is no room for the time coordinate in this framework. 

Once we have identified the ordering coordinate of $b_i$ as $x^-$, we have no choice but to identify the ordering coordinate of $R$ in eq.(\ref{rr}) as $x^-$ as well. This is quite clear,  the  operator $R$ simply shifts the charge density of the target by the charge of an extra gluon (\ref{rs}). Again, since this is a gluon in the light cone wave function, its coordinate is naturally identifiable with the longitudinal coordinate rather than the light cone time. 

In fact this interpretation goes hand in hand with the usual spatial picture of the low $x$ gluon field. One normally thinks of the valence gluons as concentrated in a "pancake"
of small extent in $x^-$, while the wee gluons are distributed in a wider longitudinal slab at smaller values of the spatial longitudinal coordinate $z$, and so at smaller values $x^-$.
Eq.(\ref{notsmall}) conforms nicely to this picture. 
In eq.(\ref{notsmall}) the values of $x^-$ in the various factors are ordered from left to write. Thus the values of $x^-$ in the factors of the charge density in the leftmost classical field $b_i$ are the smallest, the values of $\delta/\delta\rho(x^-)$ in the factor $R$ are larger and the values of $x^-$ in $\rho$'s of the rightmost $b_i$ are the largest. Following the derivation of \cite{kl} we can think of the rightmost $b_i$ and $R$ as belonging to the wave function, while the leftmost $b_i$ belonging to the conjugate wave function.
The rightmost $b_i$
in eq.(\ref{notsmall}) corresponds to the valence gluons in the wave function and thus takes up larger values of $x^-$. The wee gluons are the ones created by the shift operators $R$, and consistently with our interpretation take up smaller values of the longitudinal coordinate. From the action of $R$ eq.(\ref{rs}) it is obvious that the gluon emission probability is independent on $x^-$, which is again consistent with the common intuition that the distribution of the wee gluons is flat in rapidity.

\begin{figure}[htbp]
\begin{center}
\epsfig{file=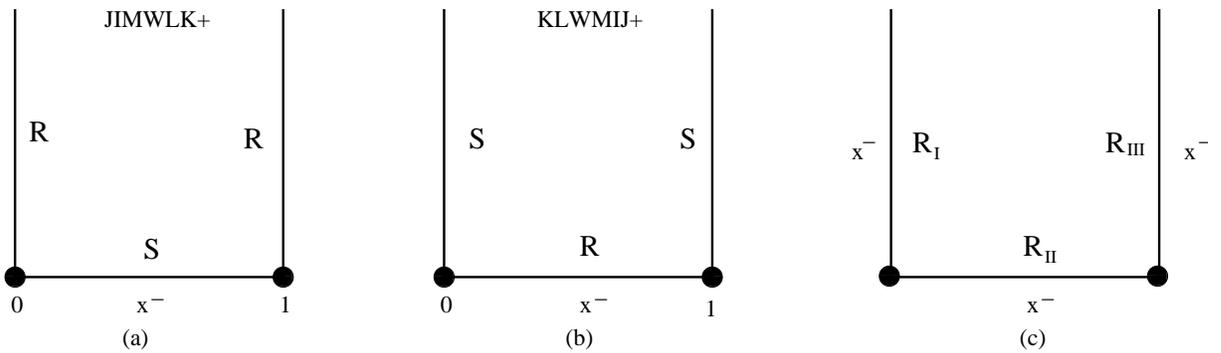,width=160mm}
\end{center}
\caption{\it Ordering structure of (a) JIMWLK+ kernel, (b) KLWMIJ+ kernel, and (c) $W$ }
\label{fig1}
\end{figure}

The weight function $W$ on which the action of the kernel is defined therefore depends on the field $\rho(x^-)$, where $x^-$ takes values in three intervals necessary for representation of both factors of $b_i$ and $R$. This contour is 
visualised in Fig. \ref{fig1}. The same structure is true for the JIMWLK+ equation if it is understood as the equation for the evolution of the target wave function in the regime of dense projectile. This is the logic adopted in Refs. \cite{kl,something,kl4}.
Of course if one writes the equation for the evolution of the projectile weight function the natural identification of the ordering coordinate there would be $x^+$.
 
The expression that is at the first sight somewhat puzzling within this interpretation, is eq.(\ref{s}), as there both $\alpha_T$ and $\rho_P$ carry the same ordering variable. However one should remember how this equation arises\cite{something}. The operator expression for the eikonal scattering matrix of the projectile charge density of the target field is
\begin{equation}
\hat\Sigma^P\,\,={\cal P}\,\,\,\exp\left\{i\int_{0}^{1} dx^-\int d^2x\,\hat\rho_P^a(x)\,\alpha_T^a(x,x^-)\right\}\label{ops}
\end{equation}
where the path ordering is along $x^-$.
The natural variable here is $x^-$, since it is the spatial variable from the point of view of the target fields and  time coordinate as far as the projectile is concerned. At this point $\hat\rho_P$ is an operator and does not carry any 
longitudinal label. To average this expression over the projectile wave function we should expand eq.(\ref{ops}) in powers of $\hat\rho_P$, endow $\rho_P$ with an ordering coordinate, which we would now call $x^+$ (as it is the spatial variable in the projectile wave function), and average with the weight function $W[\rho_P(x^+)]$. We observe however that since eq.(\ref{ops}) is path ordered with respect to $x^-$, the ordering of the operators $\hat\rho_P$ in any term in the expansion is the same as the ordering of $x^-$ coordinates in the factors $\alpha_T$ which multiply them. Thus the values of variable $x^+$ are always ordered in exactly the same way as the vaues of $x^-$. Since the correlators of $\rho_P(x_i^+)$ depend only on the ordering of $x^+_i$'s, we can simply set the value of the longitudinal coordinate of $\rho_P$ equal to that of $x^-$ in $\alpha_T$. If one wishes to find a physical picture behind this mathematical statement, the proper words are that the projectile charges and the target fields always interact at the point $t=0$, where $x^+=x^-$. Thus at the interaction point indeed the longitudinal labels of the projectile density and the target field are equal.

\subsection{Life without Wess-Zumino.}

Though the explicit form of the WZ term is known (see, e.g. \cite{kl}), it is not clear if this
knowledge can be of any use to perform the actual averages for physical observables. In this remark
we show that the following compact parameterization of the weight-functional $W$ \cite{kl4} 
\begin{equation}
W[\rho(x,x^-)]\,=\,\Sigma[R]\,\delta[\rho(x,x^-)]
\label{ansa}
\end{equation}
with $\Sigma$ - an arbitrary functional  of $R$, does not need to be augmented by the WZ term.
This form is a generalization of the form used in \cite{IM} in the framework of the dipole model with the difference that it explicitly incorporates the ordering variable, and the expression for $R$ should not be truncated at low order in derivatives as in \cite{IM}. 

The logic for this "ansatz" is the following.
Consider a hadronic state containing energetic gluons in its wave function. When scattering on any hadronic target, its $S$-matrix must be some function of the eikonal factor for the propagation.
As the $S$-matrix is given via eq.(\ref{s}) we must have
\begin{equation}
\int D\rho^a\,e^{i\int \rho^a(x,\tau)\,\alpha^a(x,\tau)}\,W[\rho]\,=\,\Sigma\left[S\right]
\label{ft}
\end{equation}
Eq.(\ref{ansa}) is specifically devised to satisfy this requirement and thus is a natural representation for a weight functional.
We will now show that it also automatically satisfies the requirement eq.(\ref{property}) and thus does not have to be augmented by a Wess-Zumino term.  
The $x^-$ ordering is relevant for $\rho$`s which are at the same transverse coordinate, thus we omit the transvers coordinate dependence in the following. 

W will use the fact that $\rho$ acts on $R$ as rotation \cite{kl4}:
\beq\label{ror}
\rho^a(x^-)\,R_{bc}(y_1^-,y_2^-)\,=\,\left[R(y_1,x^-)\,T^a\,R(x^-,y^-_2)\right]_{bc} \,\Theta(y^-_2\,-\,x^-)
\,\Theta(x^-\,-\,y^-_1)
\eeq

To illustrate the general pattern, we consider first the one and  two point functions

\begin{eqnarray}\label{1ro}
\langle \rho^a(x^-)\,\rangle&=&\int d\rho\,\rho^a(x^-)\,\,\Sigma[R]\,\delta[\rho]=\int d\rho\,\,\left[R(0,x^-)\,T^a\,R(x^-,1)\right]_{cd}
 \,\frac{\delta\Sigma[R]}{\delta R_{dc}}\,\delta[\rho]\,\nonumber \\
&=&T^a_{cd}\,\frac{\delta\Sigma[R]}{\delta R_{dc}}|_{R=1}
\end{eqnarray}

\begin{eqnarray}\label{2ro}
\langle \rho^a(x^-_1)\,\rho^b(x^-_2)\rangle&=&\int d\rho\,\rho^a(x^-_1)\,\rho^b(x^-_2)\,\,\Sigma[R]\,\delta[\rho] \nonumber \\
&=&\int d\rho\,\rho(x^-_1)\,\left[R(0,x^-_2)\,T^b\,R(x^-_2,1)\right]_{cd}
 \,\frac{\delta\Sigma[R]}{\delta R_{dc}}\,\delta[\rho]\,\nonumber \\
&=&\int d\rho\,\left\{\left[R(0,x^-_1)\,T^a\,R(x^-_1,x^-_2)\,T^b\,R(x^-_2,1)\right]_{cd}
 \,\Theta(x_2^-\,-\,x^-_1)\right.\nonumber \\
&+&\left.\left[R(0,x^-_2)\,T^b\,R(x^-_2,x^-_1)\,T^a\,R(x^-_1,1)\right]_{cd}\,\Theta(x_1^-\,-\,x^-_2) 
\right\}
 \,\frac{\delta\Sigma[R]}{\delta R_{dc}}\,\delta[\rho]\nonumber \\
&+&\left[R(0,x^-_1)\,T^b\,R(x^-_1,1) \right]_{ef}
\left[R(0,x^-_2)\,T^b\,R(x^-_2,1)\right]_{cd}
 \,\frac{\delta^2\Sigma[R]}{\delta R_{dc}\,\delta R_{fe}}\,\delta[\rho]\nonumber \\
&=&\left\{[T^a\,T^b]_{cd}\,\Theta(x_2^-\,-\,x^-_1)\,+\,[T^b\,T^a]_{cd}\,\Theta(x_1^-\,-\,x^-_2)\right\}\,
\frac{\delta\Sigma[R]}{\delta R_{dc}}|_{R=1}\nonumber \\
&+&T^a_{cd}\,T^b_{ef}\,\frac{\delta^2\Sigma[R]}{\delta R_{dc}\,\delta R_{fe}}|_{R=1}
\end{eqnarray}
The correlator is clearly piecewise constant, and depends only on the ordering of the coordinates. Moreover, the last term is symmetric under exchange of $x_1$ and $x_2$, while the interechange in the first term yields precisely the one point function, as required by eq.(\ref{property})
\begin{equation}
\langle \rho^a(x^-_1)\,\rho^b(x^-_2)\rangle=\langle \rho^a(x^-_2)\,\rho^b(x^-_1)\rangle+if^{abc}\langle\rho^c(x^-_1)\rangle{\rm sign}(x^-_2-x^-_1)
\end{equation}

Quite obviously
the same holds also for all higher correlators. The result for the three point function for example is
\begin{eqnarray}\label{3ro}
&& \langle \rho^{a_1}(x^-_1)\,\rho^{a_2}(x^-_2)\,\rho^{a_3}(x^-_3)
\rangle\,=\,\int d\rho\,\rho(x^-_1)\,\rho(x^-_2)\,\rho^c(x^-_3)\,\,\Sigma[R]\,\delta[\rho] \nonumber \\
&&=\,\left\{[T^{a_1}\,T^{a_2}\,T^{a_3}]_{cd}\,\Theta(x_2^-\,-\,x^-_1)\,\Theta(x_3^-\,-\,x^-_2)\,
+\,[T^{a_2}\,T^{a_1}\,T^{a_3}]_{cd}\,\Theta(x_1^-\,-\,x^-_2)\,\Theta(x_3^-\,-\,x^-_1)\right.\nonumber \\
&&+\,
[T^{a_1}\,T^{a_3}\,T^{a_1}]_{cd}\,\Theta(x_3^-\,-\,x^-_1)\,\Theta(x_2^-\,-\,x^-_3)\,
+\,[T^{a_3}\,T^{a_1}\,T^{a_2}]_{cd}\,\Theta(x_1^-\,-\,x^-_3)\,\Theta(x_2^-\,-\,x^-_1)\nonumber \\
&&+\,\left.
[T^{a_3}\,T^{a_2}\,T^{a_1}]_{cd}\,\Theta(x_2^-\,-\,x^-_3)\,\Theta(x_1^-\,-\,x^-_2)\,
+\,[T^{a_2}\,T^{a_3}\,T^{a_1}]_{cd}\,\Theta(x_3^-\,-\,x^-_2)\,\Theta(x_1^-\,-\,x^-_3)
\right\}\, \nonumber \\
&& \times\, \frac{\delta\Sigma[R]}{\delta R_{dc}}|_{R=1}  
\,+\, \left\{[T^{a_1}\,T^{a_2}]_{cd}\,T^{a_3}_{ef}\,\Theta(x_2^-\,-\,x^-_1)\,+\,
[T^{a_2}\,T^{a_1}]_{cd}\,T^{a_3}_{ef}\,\Theta(x_1^-\,-\,x^-_2)\,\right. \nonumber \\ 
&&+\,\left. T^{a_1}_{cd}\,[T^{a_2}\,T^{a_3}]_{ef}\,\Theta(x_3^-\,-\,x^-_2)\,+\,
T^{a_1}_{cd}\,[T^{a_3}\,T^{a_2}]_{ef}\,\Theta(x_2^-\,-\,x^-_3)
\right\}
\,\frac{\delta^2\Sigma[R]}{\delta R_{dc}\,\delta R_{fe}}|_{R=1}\nonumber \\
&&+\,T^{a_1}_{cd}\,T^{a_2}_{ef}\,T^{a_3}_{bg}\,\frac{\delta^3\Sigma[R]}{\delta R_{dc}\,\delta R_{fe}\,
\delta R_{gb}}|_{R=1}
\end{eqnarray}
Again we see explicitly that eq.(\ref{property}) is satisfied.
This illustrates the general pattern: the factors $\rho^{a_i}(x^-_i)$
in the correlator are replaced by the color matrices  $T^{a_i}$ ordered according to the ordering of $x^-$.
That is for the correlator  $\langle \rho^{a_1}(x^-_1)\,\ldots \rho^{a_i}(x^-_i)\,\rho^{a_n}(x^-_n)\rangle $ 
with  strictly order $x^-_1\,<\,x^-_2\,<\,\ldots\,<\,x^-_n$ we get
\begin{eqnarray}
\label{nro}
\langle \rho^{a_1}\,\ldots \,\rho^{a_n}\rangle \,&=&\,\sum_{k=1}^n\,\,\sum_{1\le i_1<i_2<\cdots<i_{k-1}\le n}
[T^{a_1}\,\ldots\,T^{a_{i_1}}]_{cd}[T^{a_{(i_1+1)}}\,\ldots\,T^{a_{i_2}}]_{ef}\cdots[T^{a_{(i_{k-1}+1)}}\,\ldots\,T^{a_{n}}]_{bg}\,\,\nonumber\\
&\times&
\frac{\delta^k\Sigma[R]}{\delta R_{dc}\,\ldots\,
\delta R_{gb}}|_{R=1}
\end{eqnarray}
For different orderings of $x^-_i$, the n-point function is obtained from eq.(\ref{nro}) by the rule eq.(\ref{property}).

In the dense regime the introduction of the longitudinal coordinate is not necessary \cite{JIMWLK, kl1} and the correlators must become symmetric under the interchange the factors of $\rho$. This should be seen directly from eq.(\ref{nro}). Indeed the wave function of a dense system
is characterized by a large number of gluons, say $K\gg1$. In this case $\Sigma[R]$ behaves roughly as $R^K$. The n-point function (as long as $n\ll K$)  
is dominated by the  $n$-th derivative of $\Sigma$, which as we have seen above, does not depend on $x^-$ at all:
\beq\label{nrod}
\langle \rho^{a_1}\,\ldots \,\rho^{a_n}\rangle \,\simeq\,
T^{a_1}_{cd}\,\ldots\,T^{a_n}_{bg}\,\frac{\delta^n\Sigma[R]}{\delta R_{dc}\,\ldots\,
\delta R_{gb}}|_{R=1}
\eeq

In accordance with discussion in the previous sunsection, we note that the variable $R$ in eq.(\ref{ansa}) is defined as the path ordered product along all three segments of the contour in Fig.\ref{fig1}. That is
we must understand $R$ in \eq{ansa} as the product $R=R_I R_{II}R_{III}$ (Fig. \ref{fig1},c). This does not affect the derivation of this section, and also correctly reproduces 
the correlation functions involving factors of $\rho$ that appear due to expansion of the fields $b_i[\rho]$ in KLWMIJ+, eq.(\ref{notsmall}).

\section{Self-duality and the dipole toy model.}
A simplified version of the high energy evolution - the dipole model was introduced by Al Mueller \cite{Mueller}. It describes the leading high energy behaviour in the large $N_c$ limit as long as the densities in the wave functions are not too large. As shown in \cite{kl1} the dipole evolution equation can be obtained as a well defined limit of the JIMWLK evolution.
We can define the projectile dipole creation operator $s$ and the target dipole creation
operator $r$:
\beq\label{pdipole}
s(x,y)\,\equiv\,Tr[S_F^\dagger(x)\,S_F(y)]\,;\ \ \ \ \ \ \ \ \ \ \ \ \ \ 
r(x,y)\,\equiv\,Tr[R_F^\dagger(x)\,R_F(y)]
\eeq
where $F$ indicates the fundamental representation.
The canonical annihilation operators are simply $\frac{\delta}{\delta s}$ and
$\frac{\delta}{\delta r}$ for projectile and target respectively.

If  the target weight function is a function of $s$ only, that is $W^T=W^T[s]$, the action of the JIMWLK kernel on it
in the large $N_c$ limit is equivalent to the action of the dipole kernel \cite{kl1} (see also \cite{janik}):
\beq\label{jimdipole}
\chi^{JIMWLK}\,\,W^T[s]\,=\,\chi^{dipole}_{dp}\left[s,\frac{\delta}{\delta s}\right]\,\,W^T[s]
\eeq
The dipole kernel in the dilute projectile regime can be obtained by reformulating the original Muller`s
model and it has the following form  \cite{LL1}
\beq\label{chidip}
\chi^{dipole}_{dp}\left[s,\frac{\delta}{\delta s}\right]\,=\,\bar\alpha_s\,\int_{x,y,z} K_{x,y,z}\,\,
\left[s(x,y)\,-\,s(x,z)\,s(y,z)\right]\frac{\delta}{\delta s(x,y)}
\eeq

On the other hand for dilute target which contains only dipoles in its wave function $W^T=W^T[r]$, and the action of the KLWMIJ kernel
in the large $N_c$ limit is equivalent to\cite{kl4}:
\beq\label{jimdipole1}
\chi^{KLWMIJ}\,\,W^T[r]\,=\,\chi^{dipole}_{dt}\left[r,\frac{\delta}{\delta r}\right]\,\,W^T[r]
\eeq

Just like the JIMWLK and KLWMIJ kernels, the dipole kernels in the two regimes is related by the duality transformation eq.(\ref{duality}). Since the dipole model is much simpler than the complete evolution equation, one may hope to gain some insight into some features of the "holy grail" self dual evolution by studying it. So far unfortunately there is no exact formulation of duality in the framework of the dipole model. We are unable to provide such a formulation here, instead we will consider an even further simplified toy model of the evolution, which does not involve transverse coordinates. Physically this corresponds to considering the evolution of  distribution of dipoles of fixed size at a given impact paramter.

The projectile averaged $S$ matrix $\Sigma^P$ of the toy model is given by
\beq\label{Z}
\Sigma^P\,=\,\sum_n P_n\, s^n
\eeq
Here $s$ is the scattering matrix for the scattering of a single dipole on the target and $P_n$ is the probability to find $n$ dipoles in the projectile.  This probability plays the role of the projectile weight function
\beq\label{W}
W^P(n)\,\equiv\,P_n\ \ \ .
\eeq
In the dilute projectile regime the evolution of $\Sigma^P$  is given by 
\beq\label{Z1}
\frac{d\Sigma^P}{dY}\,=\,\chi^{dipole}_{dp}\,\,\,\Sigma^P
\eeq
with the dipole kernel $\chi^{dipole}_{dp}$  
\beq\label{chidipole}
\chi^{dipole}_{dp}\left[s,\frac{d}{d s}\right]\,=\,\bar\alpha_s\,
(s\,-\,s^2)\frac{d}{d s}
\eeq
where we can think of $\frac{d}{d s}$ as of the dipole annihilation operator, while $s$ as of creation operator.
The evolution (\ref{chidipole}) is equivalent to the evolution of the probabilities $P_n$ \cite{LL1}
\beq\label{Pn}
\frac{\partial P_n}{\partial Y}\,=\,\,\bar\alpha_s\,
(n\,-\,1)\,P_{n-1}\,\,-\,\bar\alpha_s\,n\,P_n\,=\,
\,\bar\alpha_s\,(e^{-\frac{d}{d n}}\,-\,1)\,n\,P_n
\eeq
This is a simple hierarchy of equations which has an interpretation that in every step in the evolution one additional dipole is emitted with probability $\bar\alpha_s$.
Note that $n$ is a discrete variable and thus the action of $d/dn$ is not well defined. However the action of the shift operator $e^{-\frac{d}{d n}}$ on any function $F_n$ is perfectly well defined and gives
\beq
e^{-\frac{d}{d n}}\,F_n\,=\,F_{n-1}
\eeq
Thus the analog of the dipole version of the KLWMIJ equation is
\beq\label{chidilute}
\chi^{dipole}_{dt}\left(n,e^{-\,\frac{d}{d n}}\right)\,=\,\,\bar\alpha_s\,
(e^{-\frac{d}{d n}}\,-\,1)\,n
\eeq
The shift operator $e^{-\frac{d}{d n}}$
is the analog of the  ``dual Wilson line'' $R$ and creates an extra projectile dipole.

Following Ref. \cite{Mueller} we can define the dipole densities $\rho^p_k$:
\beq
\rho^p_k \,\equiv\,\frac{1}{k!}\,\frac{\delta^k}{\delta s^k}\,\Sigma^P[s]_{|_{s=1}}
\eeq 
The evolution \eq{Pn} or its equivalent \eq{chidip} leads to the equation for $\rho^p_k$ \cite{LL2}
\beq\label{rho}
\frac{d\rho^p_k}{dY} \,=\bar \alpha_s[\,k\,\rho^p_k\,+\,(k\,-\,1)\,\rho^p_{k-1}]
\eeq 

It is sometimes useful to define the amplitude of  simultaneous  interaction of $k$ dipoles on the target - $\gamma_k$. 
The $S$-matrix can be written in terms of  $\rho_k$ and $\gamma_k$  as\cite{Kovchegov} 
\beq\label{Sd2}
{\cal S}(Y)\,=\,1\,+\,\sum_k (-1)^k\,\rho^p_k\,\gamma_k
\eeq
As shown in  \cite{LL2},
the evolution \eq{rho} can be mapped onto the Balitsky`s hierarchy for the amplitudes $\gamma$:
\beq\label{gam}
\frac{d\gamma_k}{dY} \,=\bar\alpha_s\,k\,[\gamma_k\,-\,\gamma_{k+1}]
\eeq

It is instructive to see how the duality transformation works in this simple model.
It is useful to keep in mind
the following set of anologies (for the definition of $\sigma$ see below):
$$\rho\, \rightarrow \,n\,; \,\,\,\,\,\,\,\,\, \ \ \alpha\,\rightarrow\,n\,\ln\sigma\,;\ \ \ \ \ \ \  \ \ 
S\,\rightarrow\,e^{\,n\,\ln\sigma}; \,\,\,\,\,\,\,\,\,\ \ R\,\rightarrow\,e^{-\,\frac{d}{d\,n}}
$$

The $S$-matrix is computed by averaging $\Sigma^P$ over target distribution $W^T[s]$
\beq\label{Sd}
{\cal S}(Y)\,=\,\int ds\,\Sigma^P_{Y-Y_0}\,[s]\,W^T_{Y_0}[s]
\eeq
Written in this form $W^T$ gives a distribution of probabilities that a single projectile dipole scatters off the target
with the amplitude $s$. Though we have written integral over $s$, in fact $s$ is discrete, and the integral over $s$ simply should be understood as the sum over postive integers $n$ with $s=\sigma^n$. 

Let us
denote by $\sigma$ a cross section of a single projectile dipole to scatter on a single dipole of the 
target (an analog of two gluon exchange in real QCD), $\sigma=1-\gamma\alpha_s^2$ with $\gamma$ of order unity.
If we allow a projectile dipole to multiply rescatter on $m$ dipoles of the target, 
the natural model for the scattering amplitude on a $m$-dipole target state is
$s_m=\sigma^m$.  
The target by itself has a distribution of dipoles
\beq\label{Wt}
W^T(m)\,\equiv\,T_m
\eeq 
Then \eq{Sd} can be written in a manifestly symmetric form
\beq\label{Sd1}
S(Y)\,=\,\sum_{n,m} P_n(Y\,-Y_0)\,T_m(Y_0)
\,\sigma^{m\,n}\,=\,\,\sum_{n,m}\, P_n\,T_m\,e^{m\,n\,\ln \sigma}
\eeq

To derive the duality transformation in our toy model we assume that $P_n$ evolves according to the action
 of some kernel 
$\chi^{dipole}(n,\,e^{-\frac{d}{d n}})$
acting on $ P_n$:
\beq\label{Pn1}
\frac{\partial P_n}{\partial Y}\,=\,\chi^{dipole}\left(n,\,e^{-\frac{d}{d n}}\right)\,P_n
\eeq
By ``integrating by parts'' we take the kernel $\chi^{dipole\,\dagger}$ to act
on $\sigma^{n\,m}$.  The result of this integration by parts is that $d/dn\rightarrow-d/dn$, and the ordering of factors containing $n$ and $d/dn$ is reversed. We then notice that 
\beq\label{dddd}
n\,e^{\frac{d}{d n}}\, \sigma^{n\,m}\,=\,e^{\,m\,\ln \sigma}\frac{1}{\ln\sigma}\,\frac{d}{d m} \sigma^{n\,m}
\eeq

Integrating by parts  again we take the kernel to act now on $T_m$. The result of this three step procedure is the evolution equation for $T_m$
\beq\label{Pm1}
\frac{\partial T_m}{\partial Y}\,=\,\tilde\chi^{dipole\,}\left(
-\frac{1}{\ln\sigma}\,\frac{d}{d m},\,e^{\,\,m\,\ln \sigma}\right)\,T_m
\eeq
where the function $\tilde\chi$ differs from the function $\chi$ by the opposite ordering of the factors containing $n$ and $d/dn$.
In order to obtain \eq{Pm1} we used the Lorenz invariance (LI), which 
requires that ${\cal S}$ does not depend on $Y_0$. Now  the projectile-target democracy  says that $P$ and $T$ have to obey the very same
evolution law. Hence we deduce
\beq\label{dddd1}
\chi^{dipole}\left(n,\,e^{-\frac{d}{d n}}\right)\,=\,\tilde\chi^{dipole}\left(-
\frac{1}{\ln\sigma}\,\frac{d}{d n},\,e^{\,n\,\ln \sigma}\right)
\eeq
The DDD transformation can therefore be formulated as the transformation
\beq\label{dddt}
n\,\rightarrow-\,\frac{1}{ \ln\sigma}\,\frac{d}{d n} \,;\ \ \ \ \ \ \ \ \ \ \  \ 
e^{-\frac{d}{d n}}\,\rightarrow\,e^{\,\,n\,\ln \sigma}
\eeq
accomapnied by the interchange of the order of $n$- and $d/dn$-dependent factors.

We remark  that the DDD transformation (\ref{dddt}) as well as the selfduality condition (\ref{dddd1})
have been derived in a specific model for the projectile-target interaction, namely ($e^{m\,n\,\ln \sigma}$).
Although this is a reasonable ansatz, in our toy model we are not bound to choose this particular form.
Have we choosen another ansatz for $s$ the resulting transformation would be different. Thus although the existence of duality transformation follows from the Lorentz invariance and PTD, its mathematical action depends on the 
form of the projectile-target interaction.

Let us now apply the DDD transformation (\ref{dddt}) to the ``dilute'' kernel eqs.(\ref{chidipole},\ref{Pn}).
The resulting kernel is analogue of the JIMWLK equation.
\beq\label{chiden}
\chi^{dipole}_{dense}\,=\,\,\bar\alpha_s\,
\frac{1}{\ln\sigma}\,\frac{d}{d n}\left(1\,-\,e^{\,\,n\,\ln \sigma}\right)
\eeq
To make sense of the operator $\frac{d}{d n}$ we should represent it in terms of shift 
operators $\exp\{-k\,\,d/dn\}$. This reresentation is obtained using the trivial property $x=-\ln e^{-x}$:
\beq\label{dn}
\frac{d}{d n}\,=\,\sum_{k=1}^{\infty}\,\frac{1}{k}\,\left(1-e^{-\frac{d}{d n}}\,\right)^k
\eeq
The evolution equation then can be written entirely in terms of  the creation operators of $k$ dipoles $e^{-k\,\frac{d}{d n}}$  (this interpretation is in analogy with eq.(\ref{Pn})).
\begin{eqnarray}
\label{chiden11}
\chi^{dipole}_{dense}\,&=&\,\,\bar\alpha_s\,
\frac{1}{\ln\sigma}\,\sum_k\,\frac{1}{k}\,\left(1-e^{-\frac{d}{d n}}\,\right)^k\left(1\,-\,e^{\,\,n\,\ln \sigma}\right)
%&=&\,\,\bar\alpha_s\,
%\frac{1}{\ln\sigma}\,\sum_k\,\frac{1}{k}\,:\left(1-e^{-\ln\sigma}e^{-\frac{d}{d n}}\,\right)^k\left(1\,-\,e^{\,\,n\,\ln \sigma}\right):
\end{eqnarray}
%where the normal ordering in the last equality means that any of the derivatives do not act on %the functioon of $n$ in the kernel itself, but only on the function on which the kernel acts.
The expansion (\ref{dn}) has a remarkable
propery: each  term in the expansion separatly conserves probability. The subtraction of the unity
takes into account virtual contributions for $k$-dipole emission\footnote{ One should be able to obtain a similar 
representation for the JIMWLK equation by representing $\frac{\delta}{\delta \alpha}$ as an infinite series in powers of $(R-1)$.}.

The \eq{chiden11} suggests the interpretation that in the dense regime, arbitrarily many
dipoles are created during one step of evolution. 
This interpretation is indeed correct, and is consistent with the actual evolution in QCD \cite{inprep,zakopane}. However it is also clear that the kernel does not correctly account for 
the propagation of these newly born gluons through the target, as only one eikonal factor is present in eq.(\ref{chiden}). This is  an analog of the situation discussed in the beginning of the previous section. Assumption of the dense wave function of one of the colliding objects (multiple powers of $\exp\{-\frac{d}{d n}\}$) goes together with the assumption of the dilute wave function of the other 
(only first power of $\,\exp\{\,\,n\,\ln \sigma\}$). In the full kernel higher powers of 
the dipole creation operator must be also accompanied by higher powers of the eikonal scattering amplitude. The following interpretation of \eq{chiden11} looks natural. Real emission of $k$ dipoles in the full evolution kernel is accompanied by virtual corrections to all wave function components with numbers of dipoles smaller than $k$. The component of the wave function with $k$ emitted dipoles contributes to the $S$-matrix with the factor $s^k=\sigma^{nk}$, while the virtual terms contribute with lower powers of $s$, corresponding to the number of dipoles in the state whose probability is being "repaired" by the virtual correction. In the dense limit $s\ll 1$ and so the leading contribution to the $S$-matrix comes from the propagation of the state with only one emitted dipole and the state with no emitted dipoles. Thus although only one emitted dipole shows up as the "real" emission term, all the higher dipole contributions show up in the virtual correction piece: the norms of the "no extra dipole state" and "one extra dipole state" are decreased by the total probability of emitting more than one dipole in one step of the evolution. This interpretation is further strengthened by the observation that the series in the one dipole creation operator in \eq{chiden11} is an alternating sign one, which is what one naturally expects from a virtual correction.

In the dense limit where $n$ is large and $\frac{d}{d n}$ is small we can actually simplify the kernel by keeping only the first, single dipole term in the expansion (\ref{dn}).
\beq\label{chiden1}
\chi^{dipole}=\,
\bar\alpha_s\,\frac{1}{ \ln\sigma}\,\left(1-e^{-\frac{d}{d n}}\right)\,
\left(1\,-\,e^{\,\,n\,\ln \sigma}\right)
\eeq
The evolution eq. (\ref{chiden1}) possesses a remarkable property: it is selfdual under the DDD
transformation (\ref{dddt}). The kernel eq.(\ref{chiden1}) is expressed entirely in terms of the
``Wilson lines'' and in the two known limits of either dilute target or dilute projectile
reduces to the ``JIMWLK'' and ``KLWMIJ'' equations. 
The kernel \eq{chiden1} leads to the following evolution equation for the probabilities
\beq\label{Pnsd}
\frac{\partial P_n}{\partial Y}\,=\,\,{\bar\alpha_s\over|\ln\sigma|}\,\left[(1-\sigma^{n-1})\,P_{n-1}\,-
(1-\sigma^{n})\,P_{n}\right]
\eeq
This has an interesting interpretation. The coefficient in front of $P_n$ on the right hand side has the meaning of the probability per unit rapidity to emit an extra dipole from the $n$-dipole state. In this respect \eq{Pn} is clearly problematic, since for large $n$ the emission probability grows without bound. Therefore it can only be considered informative for relatively small $n$, such that $\bar\alpha_sn<1$. The behavior of \eq{Pnsd} is much better, since the emission probabilities are finite even for large $n$ and so the gluon emission probability "saturates". Since by definition $\sigma<1$, the limiting value of the probability in \eq{Pnsd} is $\alpha_s/|\ln\sigma|$. We note however that this limiting value generically is still larger than unity. Within our present model $\alpha_s$ is in principle unrelated to $\sigma$. However in QCD the analog of $\sigma$ is the dipole-dipole scattering amplitude, which behaves as $\sigma=1-\gamma\bar\alpha_s^2$ with $\gamma$ of order unity. With this parametric dependence the limiting value of the coefficient in \eq{Pnsd} is $O(1/\alpha_s)\gg1$. Thus although the emission probability given by the selfdual form of the evolution saturates, it still has the probabilistic interpretation only for $\bar\alpha_sn<1$ just like the initial expression eq.(\ref{Pn}).

The expression \eq{Pnsd} is analogous to the self dual variant of JIMWLK+ evolution recently
proposed in Ref. \cite{SMITH}. The lesson we learn from our toy model is that self duality is not sufficient to gurantee the accuracy beyond the asymptotic limits large and small density limits. In general we should expect higher powers of both $R$ and $S$ to enter the fray (see also Ref. \cite{Balitsky05} for discussion of the selfdual form of the QCD evolution).

A somewhat surprising aspect of the evolution \eq{Pnsd} is that just like eq.(\ref{Pn}) it describes only "splitting" of dipoles and does not contain "mergings" in the parlance of \cite{MSW,IT,LL3}. One could naively expect to find merging terms which lead to growth of the probability
$P_n$ proportinal to probabilities  $P_k$ with $k>n$. Such terms do not appear in eq.(\ref{dn}) nor in the self dual limit of the evolution eq.(\ref{chiden1}).
On the other hand the common wisdom is that the "splitting" evolution of the projectile dipoles is somehow equivalent to "merging" evolution in the target. The resolution of this paradox is that even though there are no merging terms in the evolution equation for the probabilities $P_n$, they do indeed apear in the equation for the $n$-dipole densities $\rho_n$ and the $n$-dipole scattering amplitudes $\gamma_n$. 
In particular
if instead of canonical dipole model \eq{Pn} we start from the selfdual evolution \eq{Pnsd} we obtain the following evolution
in terms of $\rho_k$
\beq\label{rho1}
\frac{d\rho^p_k}{dY} \,=\,\,{\bar\alpha_s\over|\ln\sigma|}\,\left[
\rho^p_{k-1}\,-\,\sum_{n=k-1}^{\infty}\rho_n^p\,\sum_{m=n+1-k}^{\infty} (\ln\sigma)^m\,C(k,m,n)\right]
\eeq
where the coefficients $C(k,m,n)$ are computable combinatorial factors. Since the expansion in $\ln\sigma$ is essentially an expansion
in $\alpha_s^2$ truncating it at  second order we obtain{\footnote{Strictly speaking this truncation is not justified. Eq.(\ref{rho1}) was derived in the approximation of large density. In this limit the higher dipole number densities carry extra powers of $\alpha_s^{-2}$ which cancel explicit powers of $\ln\sigma$ in eq.(\ref{rho1}). Thus we can only truncate the sum over $k$ but not the sum over $n$ in \eq{rho1}. The following expression therefore has to be understood within these limitations.} 
\beq\label{rho2}
\frac{d\rho^p_k}{dY} \,=\,\,\bar\alpha_s\,\left\{ k\,\rho_k^p \,[1\,+\,(k-{1\over 2})\,\ln\sigma]\,+\,(k-1)\,
\rho^p_{k-1}\,[1\,+\,(k\,-{3\over 2})\,\ln\sigma]\,+\,{k\,(k+1)\over 2}\,\ln\sigma\,\rho_{k+1}^p\right\}
\eeq
The first two terms are the same as in Balitsky`s hierarchy amended by $\alpha_s^2$ corrections.
The last term corresponds to the merging of Refs. \cite{IT,MSW,LL3}. When translated
to the equation for the amplitudes $\gamma_k$, this term leads to a contribution of the form
\beq
\frac{d\gamma_k}{dY} \,\sim\,\,\bar\alpha_s\,|\ln\sigma| \,\gamma_{k-1}
\eeq 
which is generally understood as Pomeron merging. Thus curiously enough we see that "Pomeron merging" term in the evolution of the amplitudes $\gamma_k$, when translated into the language of probabilities $P_n$ is in fact no different from dipole splittings. Physically this is perfectly reasonable, as boosting the hadronic state can never lead to the decrease in number of gluons or dipoles in the wave function. These numbers always grow, but the rate of their growth decreases at high energy. The Pomeron mergings describe the decrease in the rate of growth of the cross section, and not the decrease in the number of partons.

\section*{Aknowledgments}

We thank Ian Balitsky, Yoshi Hatta, Genya Levin, Larry McLerran and Urs Wiedemann for discussions related to the subject of this paper.

\end{document}